# Phase Trombone Design in the Beam Transfer line for the Project PIP-II at Fermilab *

M. Xiao[†], F. Ostiguy and D. Johnson

Fermilab, Batavia, USA

*Abstract*

PIP-II beam transfer line (BTL) to transport the beam from PIP-II Linac to the Booster ring at Fermilab. One crucial aspect of the BTL design involved the implementation of collimators. These collimators play a vital role in removing large amplitude particles from the beamline that might otherwise miss the horizontal and vertical edge of the foil at the point of the Booster injection. To ensure the effectiveness of these collimators, simulation was conducted to determine their optical placement within the BTL. The simulation revealed that the precise control of accumulated phase advances between the collimators and the foil was crucial. To achieve this control, a phase trombone is needed within the BTL. The phase trombone serves as an adjustment mechanism, allowing for the fine-tuning of the phase advance between the collimators and the foil, thus optimizing the performance of the collimation process. This paper discusses and presents the details of the design and implementation of the phase trombone within the PIP-II BTL.

## INTRODUCTION

The Proton Improvement Plan II, or PIP-II, is an essential enhancement to the Fermilab accelerator complex, powering the world's most intense high-energy neutrino beam on its journey from Illinois to the Deep Underground Neutrino Experiment in South Dakota – 1,300 kilometres (800 miles) [1]. PIP-II comprises an 800-million-electron-volt linear accelerator, or linac, based on superconducting radio-frequency technologies. The beam that emerges from the PIP-II linac will be directed to the Fermilab accelerator complex.

A portion of the beam will be injected into the existing Booster ring, which will strip the two electrons from the H-minus ions and then accelerate the bare protons to 8-billion-electronvolts, or 8 GeV. Figure 1 presents the layout of the PIP-II Linac complex.

The PIP-II Beam Transfer Line (BTL) transport the beam from the PIP-II Linac to the Booster ring. The lattice design of the BTL is final [2], and the 3D CAD model is shown in Figure 2. A FODO lattice is used for the BTL. Four achromatic ARCs, each with 4 FODO cells, connected by a straight section with 8 FODO cells. Phase advances of the FODO cell is 90° in both horizontal and vertical planes. In the region of the Injection Girder of the Booster, the design meets the requirement of the civil constructions and also accommodate the constrains to avoid the interference between the BTL and the Booster ring. The roll angle of the dipoles in ARC2 in the previous design has been eliminated, the lattice now is flat. There will be 2 sets of the switch magnet and septum, one for the BAL (Beam Absorber Line) and one for the Mu2e line for the future upgrade.

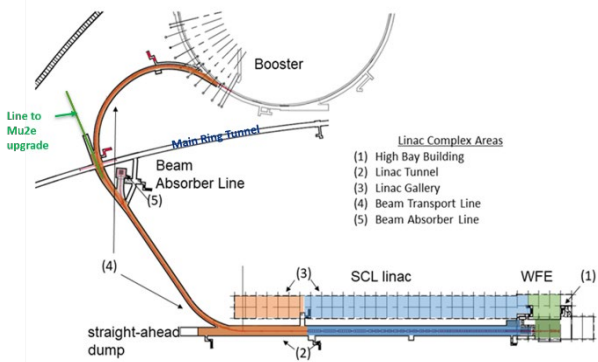

Figure 1: Layout of the PIP-II Linac Complex

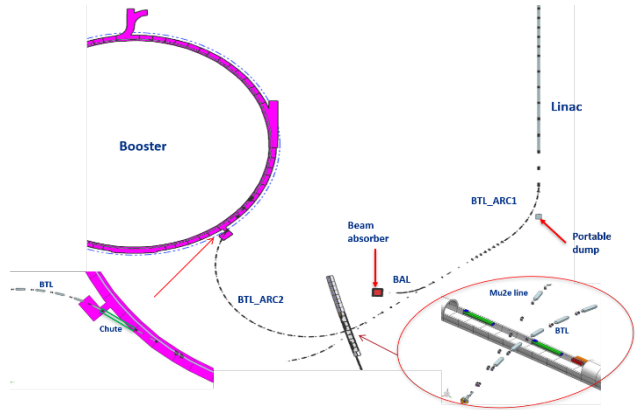

Figure 2: Layout of 3DCAD Model of the PIP-II complex.

## COLLIMATIONS IN THE BTL

One crucial aspect of the BTL design involved the implementation of collimators. These collimators play a vital role in removing large amplitude particles from the beamline that might otherwise miss the horizontal and vertical edge of the foil at the point of the Booster injection. To minimize the number of parasitic hits, the beam must be positioned as close as possible to the lower corner of the injection stripping foil. The minimal achievable distance is determined by the competing need to limit losses due to particles populating the distribution tails missing the foil.

___________________
* Work supported by Fermi Research Alliance, LLC under Contract No. De-AC02-07CH11359 with the United States Department of Energy †meiqin@fnal.gov.

The overall function of the collimation system is to collimate in a controlled manner the far tails of the beam transverse spatial distribution so as to obtain a sharp rectangular edge at the foil, illustrated in Figure 3.

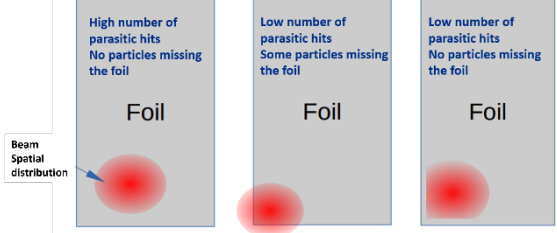

Figure 3: 3 cases of beam distribution at foil

To ensure the effectiveness of these collimators, simulation was conducted to determine their optical placement within the BTL. One of the most important constrains is that the accumulated phase advance between collimator and foil should be as close as possible to an integer multiple of π. Since collimation can be performed on either side (L/R or T/B), collimator locations that are (almost) nπ apart are (almost) equivalent. In the BTL, the total accumulated phase advance between collimator H/V locations and foil are not exact multiples of π. This phase advance depends on a number of factors: (1) Accounting for the space needed for a vacuum connection and the enclosure assembly, the primary collimator jaw must be positioned at least 650 mm (25 in) downstream of a quadrupole. (2) The phase advance per cell both within the BTL and in the arcs is very nearly but not exactly π/2 (due to bending magnet edge focusing and space charge). These deviations from π/2 accumulate over many cells. (3) The H/V phase advances from the last cell of the 2nd arc to the foil are constrained by matching requirements for the Booster painting scheme.

The best candidate locations are the 8 dispersion-free FODO cells (16 half-cells) in the straight section between the two arcs. The quad-to-quad separation in each half-cell is approximately 5.5 m. Many of the straight section half-cells are claimed for instrumentation or for future bunch cavities. After careful consideration of other needs and requirements, it was determined Cell 1 was selected for vertical collimation, Cell 2 for horizontal collimation. Collimator jaws are located approximately 635 mm (~ 25 in) downstream of the nearest upstream quadrupole。The simulation result of the beam distribution with collimators are shown in Figure 4. The particle distribution used is a "realistic" (and probably pessimistic) distribution that has a somewhat larger emittances and dp/p than the nominal specification. It was obtained by start-to-end tracking in a linac with errors.

## PHASE ADVANCE CORRECTION

Figure 4 shows that we are not able to obtain a sharp, clean spatial distribution edge at the foil with both Horizontal and vertical collimators, although the main function of the BTL collimators is to sharpen the edges of the transverse spatial distribution to allow the beam to be positioned as close as possible to the foil corner. The reason is that the accumulated phase advance determines the orientation of the collimator cut in the phase space plane at the foil location, but in the BTL base lattice, the accumulated phase advances between the collimators and the foil is not exactly an integer multiple of π. If the accumulated phase advances can be adjusted, a near ideal edge orientation in phase space can be achieved, as shown in Figure 5

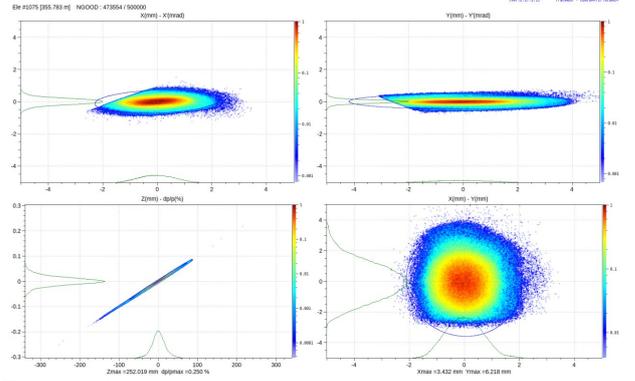

Figure 4 beam distribution with vertical collimator at cell 1 (2nd half cell) and Horizontal collimator at cell 2 )1st half cell). No phase advance correction.

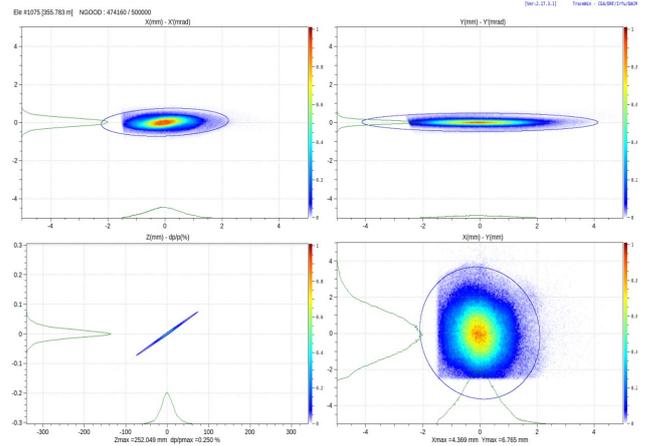

Figure 5 Distribution at foil after phase advance correction applied: -30° in Horizontal and +30° in vertical plane

## PHASE TROMBONE DESIGN

There are 8 FODO cells in the BTL straight section lattice, shown in Figure 6 are the beta-functions (solid line - βx, dotted line – βy). Highlighted 3 FODO cells (cells 6,7,8) will be redesigned to function as a phase trombone. 6 individually powered quadrupoles allow the lattice functions to remain matched to their nominal design values at both upstream and downstream boundaries (cell 5 2nd half and 2nd arc entrance). Within the trombone region the β-functions may vary, to allow adjusting the phase advance accumulated between each collimator location and the foil. Two schemes, symmetry structure and broken symmetry structure, were investigated. It was found that the symmetry structure gives more even beta-functions and the strength of each quad requested are within the maximum current of the power supply. Figure 7a presents the beta-functions of the trombone when the phase advance adjustment is set to $d\mu_x$=-30° and $d\mu_y$ =-60°. Meanwhile, Figure

7b provides an overview of the entire BTL lattice. Notably, the lattice outside the trombone section remains unchanged. The adjustable region for the phase advance in this phase trombone section is (±90º, ±90º), in both horizontal and vertical planes. Table 1 listed the strength of the 6 quadruploes used and maximum beta-functions in the trombone section for various cases.

Table1 phase advance changes $(d\mu_x, d\mu_y)$ vs quads strengths, and the maximum $\beta x$ and $\beta y$

| Adjusted phase/ parameters | $(d\mu_x, d\mu_y)$ (0,0) (Nominal) | $(d\mu_x, d\mu_y)$ (0,0) | $(d\mu_x, d\mu_y)$ (20º, 20º) | $(d\mu_x, d\mu_y)$ (-20º, 20º) | $(d\mu_x, d\mu_y)$ (20º, -20º) | $(d\mu_x, d\mu_y)$ (0, 20º) | $(d\mu_x, d\mu_y)$ (20º, 0) | $(d\mu_x, d\mu_y)$ (32º, 32º) |
|---|---|---|---|---|---|---|---|---|
| $K_{QF6}$ (m$^{-2}$) | 1.209 | 1.209 | 1.248 | 1.153 | 1.234 | 1.056 | 1.154 | 1.273 |
| $K_{QD6}$ | -1.187 | -1.221 | -1.227 | -1.202 | -1.160 | -1.138 | -1.220 | -1.259 |
| $K_{QF7}$ | 1.209 | 1.192 | 1.305 | 1.130 | 1.278 | 1.195 | 1.115 | 1.347 |
| $K_{QD7}$ | -1.187 | -1.186 | -1.293 | -1.266 | -1.117 | -1.112 | -1.171 | -1.338 |
| $K_{QF8}$ | 1.209 | 1.214 | 1.281 | 1.195 | 1.268 | 1.364 | 1.176 | 1.307 |
| $K_{QD8}$ | -1.187 | -1.150 | -1.247 | -1.241 | -1.138 | -1.137 | -1.136 | -1.264 |
| $(\beta_x)_{max}$ (m) | 19.969 | 20.441 | 21.137 | 20.866 | 19.982 | 25.000 | 21.166 | 22.431 |
| $(\beta_y)_{max}$ (m) | 20.268 | 21.144 | 21.123 | 21.125 | 21.146 | 22.972 | 21.147 | 21.588 |

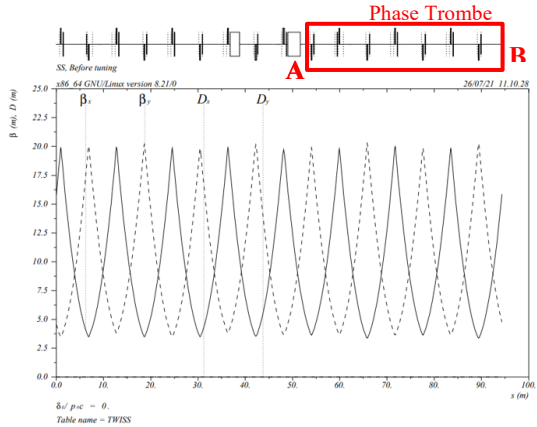

Figure 6: 8 FODO cells in the BTL lattice.

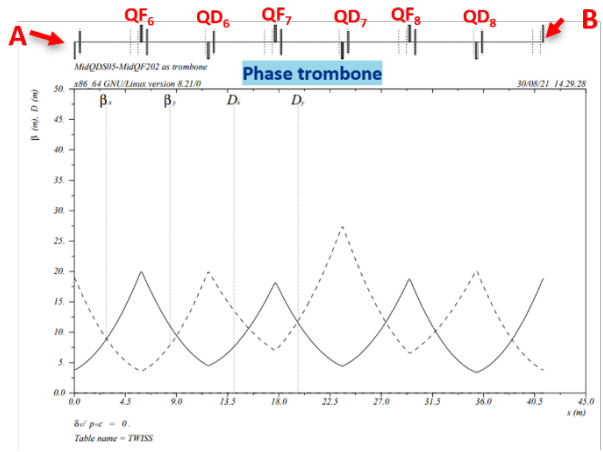

Figure 7a: Trombone in the BTL straight section; the phase advance adjustment of $d\mu_x$=-30º, $d\mu_y$=-60º.

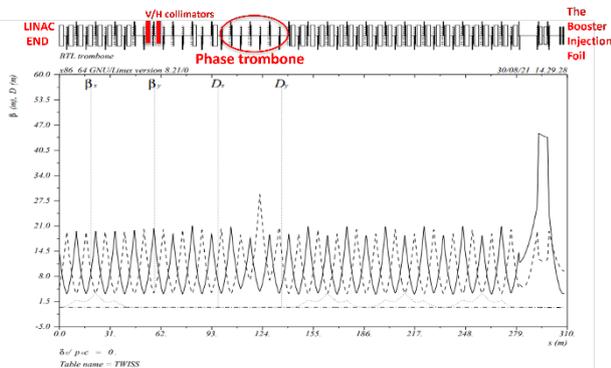

Figure 7b: Lattice of the BTL. the phase advance adjustment of $d\mu_x$=-30º, $d\mu_y$=-60º.

## CONCLUSION

A phase trombone is designed in the BTL lattice. The adjustable region of the phase advance of this phase trombone section is (±90º, ±90º), both in horizontal and vertical planes. It is used to adjust the accumulated phase advances between the collimators (both horizontal and vertical) and the foil at the Booster injection, to as close as possible to an integer multiple of π. A near ideal sharp edge orientation in phase space can be achieved without perturbing the lattice functions at the foil.